# Replacing the computer mouse

Franck Dernoncourt
francky@mit.edu
http://francky.me

August, 2012

*Abstract*—In a few months the computer mouse will be half-a-century-old. It is known to have many drawbacks, the main ones being: loss of productivity due to constant switching between keyboard and mouse, and health issues such as RSI. Like the keyboard, it is an unnatural human-computer interface. However the vast majority of computer users still use computer mice nowadays.

In this article, we explore computer mouse alternatives. Our research shows that moving the mouse cursor can be done efficiently with camera-based head tracking system such as the SmartNav device, and mouse clicks can be emulated in many complementary ways. We believe that computer users can increase their productivity and improve their long-term health by using these alternatives.

## I. INTRODUCTION

As defined by Wikipedia, a computer mouse is a pointing device[1], i.e. a human interface device that allows a user to input spatial (continuous, and multi-dimensional) data to a computer, that works by detecting two-dimensional motions relative to its supporting surface. Physically, a mouse consists of an object held under one of the user's hands, with one or more buttons. Upon receiving those inputs, the computer's operating system moves the mouse pointer (a.k.a. mouse cursor) on the screen, which tries to match the mouse's movements as accurately as possible.

The first mouse prototype was built in 1964 by Douglas Engelbart at the Stanford Research Institute[2]. We will therefore soon celebrate its 50-year anniversary.

While computer mice are very useful, they still have important drawbacks. The first one is productivity: constantly switching between the mouse and the keyboard is inefficient. For that reason, many Linux advanced users strongly rely on shortcuts to avoid using the mouse. This solution is hardly accessible for Microsoft Windows users and requires both computer skills and a time commitment that the vast majority of computer users don't have.

The second drawback is health-related: the strain using a computer mouse puts on hands and forearms may cause repetitive stress injuries (RSI), which includes many pathological conditions[3]. Billions of dollars are spent worldwide annually as a consequence of RSI [1]. See Appendix A for more information on RSI. Obviously, people with a broken, paralyzed or amputated arm may also have a hard time using a computer mouse.

Our main motivation is therefore twofold: productivity and health, which are highly interdependent. We may add a third one: getting closer to natural user interfaces, which is intertwined with the two others. Natural user interfaces are interfaces that aim to be invisible to the user, such as some speech interfaces[4] : see Appendix B for more information on natural user interfaces.

We do not aim at replacing the mouse as a pointing device but instead replacing the mouse as a piece of hardware. We therefore stay in the mouse paradigm and focus on the hardware interface to avoid having to use the hand to move the pointer. The computer mouse allows two kinds of action: moving the mouse cursor and sending mouse clicks. We need to find solutions (software or hardware) to perform these two actions.

In the first section of the article we analyze solutions that enable the user to move the mouse cursor and in the second section we focus on emulating mouse clicks.

Hands-free mice are numerous:

- Camera based head tracking systems: SmartNav, Tracker Pro, FreeTrack, HeadMouse Extreme and HeadMaster,
- Mouth-operated joystick types: the TetraMouse, the QuadJoy, the Jouse2, the IntegraMouse,
- Footmice: BiLiPro, Flip Flop Mouse, Footime Foot ControlledMouse,
- Brain-computer interaction: the Emotiv EPOC neuro-headset, the NeuroSky MindSet/MindWave,
- Eye tracking.

However, some of them only allow one of the two the above-mentioned actions, i.e. moving or clicking. The point of this article is to select and present the best devices among this plethora of solutions.

---

[1] http://en.wikipedia.org/wiki/Pointing_device

[2] See http://en.wikipedia.org/wiki/Mouse_(computing) and http://www.dougengelbart.org/firsts/mouse.html for more information on computer mouse history.

[3] such as Adhesive Capsulitis (Frozen Shoulder), Bursitis, Carpal Tunnel Syndrome, Cramp of the Hand (Writers' Cramp), Cubital Tunnel Syndrome, De Quervain's Syndrome, Dupuytren's Contracture, Epicondylitis (tennis / golfer's elbow), Ganglion Cyst, Peritendinitis, Rotator Cuff Syndrome, Tendinitis, Tenosynovitis, Trigger Finger / Thumb, and Vibration-induced White Finger (most of this list comes from http://www.repetitivestraininjury.org.uk/causes-of-rsi.html).

[4] http://en.wikipedia.org/wiki/Natural_language_user_interface

## II. MOVING THE MOUSE CURSOR

First we quickly review all the devices that allow to move the mouse cursor. Then in the second subsection we present one of the best solutions we have found so far, namely the camera-based head tracking system SmartNav. We finish this section by some considerations on eye tracking.

### A. Review of all the devices

Among the camera-based head tracking systems, SmartNav is the cheapest (300-400 USD), and it supports Windows or Mac (add 200 USD for the Mac software to make it compatible with SmartNav) but not Linux. The alternatives are TrackerPro[5] and HeadMouse Extreme[6]: they cost around 1,000 USD and work on Windows, Linux and Mac OS X. The main difference between those devices is that the firmwares of the latter truly emulate a mouse[7], while SmartNav necessitates to install a program.

Unlike camera-based head tracking systems, mouse-operated joysticks are pretty intrusive since one has to put them in the mouth, but this solution is interesting for people who cannot move their head or have severe pain in the neck. The TetraMouse is the cheapest by far and seems to be at least as good as the others. However, due to budget constraints, we have not tried it personally.

Footmice and eye tracking systems are less precise than camera-based head tracking systems. Furthermore footmice might cause stress on the feet, ankles or the legs.

Brain computer interaction are mostly useless to move the mouse cursor given the current technologies. Note that the Emotiv EPOC neuroheadset contains a gyroscope (device for measuring or maintaining orientation), thanks to which the user can move the mouse cursor as precisely as a computer mouse. However, wearing the Emotiv EPOC neuroheadset for a long period is not comfortable at all: from that perspective, camera-based head tracking systems is arguably better.

### B. SmartNav

SmartNav uses an infrared (IR) camera to track head movements, as shown in Figure 1. The user reflects IR light back to the SmartNav, which sends instructions to the computer to move the mouse cursor. SmartNav tracks reflections from a tiny dot, which the user can place anywhere, typically the forehead. As a result, it only takes a few seconds to enable the device. SmartNav costs between 300 and 400 USD, which is cheap compared to many hands-free mice. It supports multiscreen configuration (tested with 6 monitors, with total

---

[5]http://www.ablenetinc.com/Assistive-Technology/Computer-Access/TrackerPro

[6]http://www.orin.com/access/headmouse/

[7]i.e. there is no need to install any program on the computer put aside the universal human interface device drivers, which are normally already installed on the OS. This is actually a huge upside as it ensures the compatibility with any operating system that can be used with a traditional computing mouse, and the pointer controlled by the camera-based head tracking system will lag just as much as a mouse's pointer, but not more. In our tests with Windows 7 SP1 x64 Ultimate, installing/running SmartNav with administrator privileges, and running SmartNav with its priority set to high, SmartNav's pointer did not lag more than a mouse, but we cannot ensure this for any configuration.

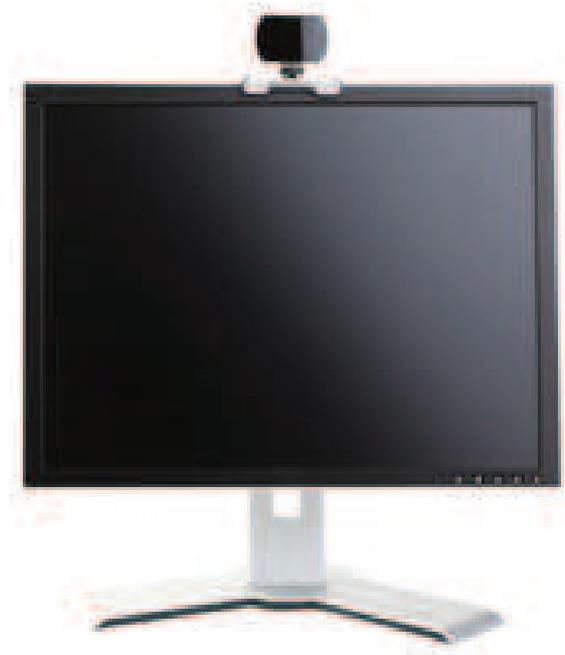

Fig. 1. SmartNav's IR camera is typically positioned on the top of the screen. Source: SmartNav's website.

resolution of 5520x1848) and works on Windows and Mac OS (but not on Linux).

The SmartNav official page[8] gives more information, and here is an excerpt:

*How does the technology work? Infrared light is emitted from the camera's LEDs and is reflected back to the imager by a corner cube reflector (3M safety material). This reflected light is imaged by a CMOS sensor and the video signal is passed to the preprocessing electronics. The video signal is thresholded against a reference level and all passing data is sent to the USB microcontroller to send to the PC for object tracking. In order to increase the signal to noise ratio an IR filter that passes only 800nm and above is placed between the imager lens and the outside world. The SmartNav can image any IR source; typically this is reflective material or an active IR source such as an LED. A user may track many different objects by placing reflective dots or LEDs on the object. The SmartNav has a 45 degree field of view and anything being tracked must stay in that field of view.*

*How are head movements tracked? The user places a tiny reflective dot on the part of the body he wants to control the cursor with. Preferred options include: head, hand, hat, glasses, microphone. The user can also make his own reflective marker with NaturalPoint's tracking material.*

*Where does the user put the SmartNav? SmartNav mounts on top of the monitor, laptop or communication device facing the user. SmartNav can also be threaded onto a mini tripod and sit next to the computer. The device can be placed anywhere*

---

[8]http://www.naturalpoint.com/smartnav/products/about.html

*as long as it can see the reflective accessory the user have chosen to wear.*

*How much do the user move? Less than an inch of head movement is more than enough to move the cursor across the entire screen. This is also adjustable in the software settings. SmartNav has a 45 degree field of view, and usually sits about 2 feet away from your head. Thus the user has almost two feet of free "head space" in which to move that simple inch.*

*C. Eye tracking*

Eye tracking systems could provide an interesting solution for moving the mouse in the future. Current systems are either less precise than camera-based head tracking systems (e.g. ITU GazeGroup) or extremely expensive (e.g. Tobii is over 10 k€ and EyeTech is over 6 kUSD). Furthermore, they often do not support multiscreen. Theoretically they only require a webcam, but in practice to achieve a decent precision they may need several webcams and/or an IR camera. However, we hope that in the long-run eye tracking devices will turn out to be cheap and accurate. The latest results look promising, e.g. [2]. An overview of eye tracking systems can be found in [3], in which Figure 9 (reproduced in this article as Figure 8 in Appendix) lists the accuracy for many gaze estimation methods.

## III. EMULATING MOUSE CLICKS

There exist many ways to emulate mouse clicks. Here are the main categories:

- Hotkeys: re-map keys from the keyboard and assign them to emulate the left, right and middle mouse buttons.
- Dwell clicking software: watch as the user moves the mouse cursor. When the cursor stops moving for a pre-determined amount of time (usually around 1 second), the dwell clicking software will send a mouse click. The user can have the software send left clicks, right clicks or double clicks. Figure 2 shows a dwell clicking application.
- Footswitches: allow the user to send mouse clicks by pushing a pedal.
- Speech recognition: set a few voice commands that the user can say to emulate mouse clicks.
- Facial expression recognition: maps facial expressions such as eye blink, wink or smile to mouse clicks.
- Brain-computer interaction: maps concepts to mouse clicks. When the user thinks of one concept, a mouse click is sent.

Firstly we emphasize on the complementarity of these devices: unlike moving the mouse cursor, emulating mouse clicks can be done pretty efficiently with many devices, but none of them are perfect, hence the idea to combine them to take advantage of their diversity. Then we focus on facial expression recognition.

*A. Complementarity of the devices*

Each of the solutions have pros and cons, as summarized in Table I. We personally advise one of those combinations:

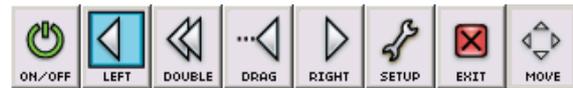

Fig. 2. SmartNav's dwell clicking software (free). The user needs to hover on the button corresponding to the mouse type he wants to. No keyboard is required for any action: everything is done by moving the mouse cursor.

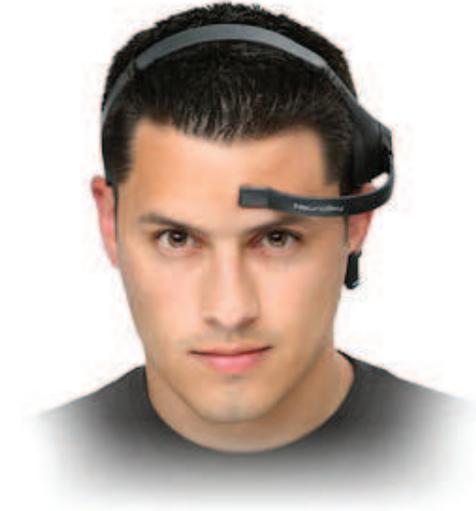

Fig. 3. The NeuroSky MindWave headset. Source: NeuroSky website.

- Hotkeys alone
- Dwell clicking alone
- Dwell clicking + (hotkeys and/or footswitches and/or speech recognition)

Note that while hotkeys and speech recognition are straightforward to use, there is a small learning curve for dwell clicking. Getting fully at-ease with it often requires a few days, however it is more efficient than what one might believe at the first sight. Facial expression recognition might be useful but neither very accurate nor comfortable. The same inconveniences are present in brain-computer interaction, which is even less accurate and has a much higher latency. This is bound to change in the future though.

*B. Facial expression recognition*

Facial expression recognition or brain-computer interaction requires a headset. This might sound surprising for facial expression recognition: in fact, a webcam could theoretically suffice but the current applications on the market do not perform well enough to be of pragmatic use. Developers of headsets for brain-computer interaction have noticed on their side that much noise comes from the movements of facial muscles: they subsequently added facial expression recognition features on their headsets.

The two most popular neuroheadsets, which can do both facial expression recognition and real brain-computer interac-tion, are: the NeuroSky MindWave headset (Figure 3) and the Emotiv EPOC neuroheadset (Figure 4).

TABLE I
COMPARISON OF SOLUTIONS TO EMULATE MOUSE CLICKS

| Solution | Pros | Cons |
| --- | --- | --- |
| Hotkeys | Free, easy to use, no latency | Requires to use the keyboard, not so good for RSI (much better than mouse clicks though) |
| Dwell clicking | Free, easy to use, no latency, hands-free | Requires to wait ∼1 second before click is sent, takes some time (∼1s) to switch between mouse click types |
| Speech recognition | Easy to use, hands-free, already integrated within Dragon NaturallySpeaking | High latency (∼1s), put some strain on the voice, can be inconvenient to use in open spaces |
| Facial expression | Hands-free | Quite expensive (100-300 USD), detection is not 100% accurate, headsets are not comfortable |
| Brain-computer interaction | Hands-free | Latency, quite expensive (100-300 USD), not accurate enough to emulate mouse clicks, headsets are not comfortable |

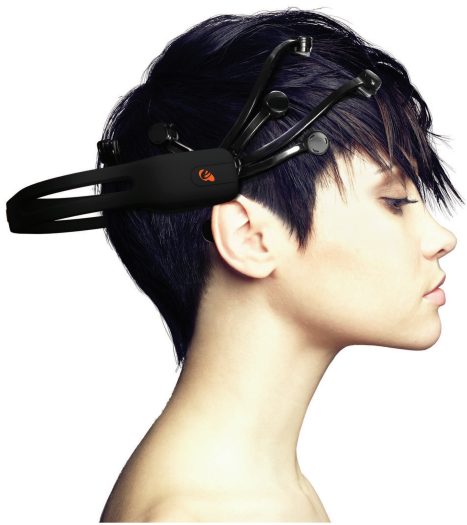

Fig. 4. The Emotiv EPOC neuroheadset. Source: Emotiv website.

The NeuroSky MindWave has one sensor and provides only three values: attention, meditation and eye blinking[9]. Only the latter is of interest to us as the first two ones are useless for computer interaction. As there was no application to map the eye blink to the mouse click, we wrote a program we called NeuroClick for that purpose. The source code (C) as well as the binaries can be found on http://francky.me/software.php#Neuroclick2012. The eye blink detection accuracy is approximately 90% accurate, due to the NeuroSky MindWave API which misses around 10% of eye blinks. If one does not want to click but needs to blink, it should not be too annoying since most of the time the mouse cursor is placed on a neutral position so it does not matter if a click is sent. Furthermore, for Neurosky to maximize the precision of the eye blink detection one has to blink strongly, which means that "natural" blinks are less likely to trigger an unwanted blink. However, as one commentator asked, *"What would happen to a conventional mouse that only recognized 90% of your left clicks and has no right clicks? Your answer to that question is also the reason why the Mindwave needs a lot of further development."*. In addition to that, the NeuroSky MindWave is highly uncomfortable: as shown in Figure 7 in Appendix, most NeuroSky users cannot comfortably wear the headset for more than one hour in a row.

The Emotiv EPOC neuroheadset can be seen as a more advanced headset than NeuroSky MindWave: it has 14 sensors, the applications provided offer more features and mapping the eye blink to a mouse click or a key is straightforward. Facial expression recognition is more accurate and can detect more expressions than simply eye blink, such as wink and smile. However, it is three times more expensive than the NeuroSky MindWave: 300 USD for the Emotiv EPOC vs. 100 USD for the MindWave. Moreover, it has a couple of inconveniences: its sensors are wet, and it is very long to place (a few minutes). Also, like NeuroSky users, most Emotiv users cannot comfortably wear the headset during more than one hour in a row (see Figure 9 in Appendix). As a side note, the Emotiv EPOC neuroheadset has a gyroscope that allows the user to move the mouse cursor as precisely as with a camera-based head tracking system.

## IV. GETTING RID OF THE KEYBOARD TOO

When one wants to replace the computer mouse one might also wants to rethink one's use of the keyboard. We briefly present the most accurate speech recognition software available for Windows and Mac OS X: Dragon NaturallySpeaking. In one study with average computer users, the average rate for transcription was 33 words per minute, and 19 words per minute for composition [4]. An average professional typist types usually in speeds of 50 to 80 words per minute[10]. Using speech recognition, one can easily achieve over 100 words per minute with more than 95% accuracy.

---
[9]More precisely, the NeuroSky MindWave provides eight brain waves from which the API computes attention, meditation and eye blinking.

As stated in Dragon NaturallySpeaking 11 User Guide: *One reason to use Dragon is to boost your productivity. Another is to reduce the strain using a computer puts on hands, eyes, shoulders, etc. Maybe you like the idea of being able to lean back in your chair, put your feet up on the desk, and still get work done.*

This is obviously a commercial statement, but one that is true. Beyond the speech recognition accuracy, one can add words to the vocabulary. Many voice commands are available, such as any shortcut (e.g. copy paste), typing a predefined text, switching windows, browsing the web, sending e-mail and launching programs. Custom commands can also be easily defined, as illustrated by Figure 12. See Appendix F for more information on Dragon NaturallySpeaking. Unfortunately the program is not available on Linux.

## V. Discussion and Conclusion

By way of conclusion, to our knowledge, camera-based head tracking systems such as SmartNav are currently the best solution to move the mouse cursor and dwell clicking combined with (hotkeys and/or footswitches and/or speech recognition) is the best solution to emulate mouse clicks. SmartNav is as efficient as a computer mouse for most computer uses (e.g. gaming or graphics editing are exceptions). In the near future we expect that eye tracking and facial expression recognition will be cheaper, even possibly free and open source, while being as accurate as a computer mouse. In a further future, brain-computer interaction is likely to become a serious competitor for both mouse cursor moving and mouse clicks.

## Acknowledgment

In alphabetical order:
- Thibault Carron
- Matthieu Cisel
- François De Forges
- David Dernoncourt
- Loïc Février
- Olivier Filipowicz
- Geoffrey Garcia
- Armand Iranpour
- Jean-Marc Labat
- Jenny Lee
- Sam Neurohack
- The Quora community

---

[10]More statistics on http://en.wikipedia.org/wiki/Words_per_minute

APPENDIX

A. Repetitive stress injuries RSI

*1) What are the causes of RSI?:*
From the *Ayurveda Clinics*[11]:
RSI arises due to the following factors:
- Prolonged repetitive, forceful, or awkward hand movements
- Poor posture
- Static loading or holding a posture which promotes muscle tension for a long period
- Poor conditioning of the heart and lungs, and poor muscle endurance
- Direct mechanical pressure on tissues
- Cold work environment
- Poorly fitting furniture
- Basic inadequacies of keyboard, monitor and workstation design
- Work organisational and psychosocial issues

*Restricted blood flow is often the culprit. Lack of blood to the muscles, tendons and nerves can cause or aggravate a host of conditions, even, perhaps, arthritis. When you tense a muscle to just 50% of its ability, the blood flowing through the capillaries in the muscle can be completely shut off. Tensed muscle fibres pressure the capillaries thereby restricting the blood flow. As the muscle is continually tensed and no fresh blood is supplied, it switches from aerobic (with oxygen) to anaerobic (without oxygen) metabolism. This produces by-products such as lactic acid which can build up and cause cell damage and pain. Subsequently, the neighbouring muscles work harder to help carry the load, but because they are not designed to do the job as efficiently, those muscles fatigue (anaerobic) even faster.*

*Muscle tension, therefore, restricts blood flow and restricted blood flow causes more tension in other muscles. If the muscles are not allowed to relax, cellular degeneration can rapidly increase as a vicious cycle takes hold. The tensed muscles also pressure surrounding nerves which causes tingling, numbness, and more subsequent injury. In addition, the lack of blood increases the likelihood of degeneration and inflammation throughout the system and, of course, retards healing. And though the cycle may stop when you rest your hands, by the time you feel any symptoms, the damage has already started. Consequently, it will take less stress to bring on symptoms in the future.*

*Repeated tensing of the hand can cause the fibres of the tendons running through the carpal tunnel to separate or break. This causes friction between the tendon and its sheath (tenosynovium) and ultimately tendonitis. Tenosynovitis occurs when the sheath cannot properly lubricate the tendon it surrounds due to the repetitive hand movement and the sheath itself becomes inflamed. Tightly gripping something for too long and forceful movements can lead to problems as well. The two most common forms of RSI are Carpal Tunnel Syndrome and Tendon Injuries*

[11] http://www.ayurvedaclinics.in/Repetitive-Stress.html

*2) Useful links on RSI:*
- http://www.rsi.org.uk/pdf/ULDs_Overview.pdf
- http://rsi.org.uk/text_only/whatis/prevalence.html
- https://en.wikipedia.org/wiki/Repetitive_strain_injury
- http://web.mit.edu/atic/www/rsi/
- http://www.scriven.com/RSI/rsiadvice.html

B. Natural User Interfaces (NUI)

*1) What are natural user interfaces:* As described by Wikipedia, a natural user interface, or NUI, loosely refers to a user interface that is (1) effectively invisible, or becomes invisible with successive learned interactions, to its users, and (2) is based on nature or natural elements (i.e. physics). Figure 5 shows the evolution of user interfaces. Many solutions we have described in his article to replace the mouse such as SmartNAV or speech recognition tend toward natural user interfaces. This is therefore an interesting field to look into when trying to replace conventional computer devices (mouse, keyboard, traditional screen, ...). For example, LeapMotion, a device that recognizes hand movements and that is announced to be released in early 2013, brands itself as natural user interface which could revolutionize the mouse: *This is like day one of the mouse. Except, no one needs an instruction manual for their hands.*[12]. Microsoft Kinect is a motion sensing input device that can be connected to Windows PCs and reflects Microsoft's interest in natural interfaces. Wired gloves are input devices for human-computer interaction worn like a glove. For example, the Peregrine gloves can perform over 30 unique actions by touching fingers. Even lasers can be used as naturals interfaces, as demonstrated by hacker Hector Martin[13].

*2) Interesting links on natural user interfaces:*
- http://en.wikipedia.org/wiki/Natural_user_interface
- http://research.microsoft.com/en-us/collaboration/focus/nui/
- http://leapmotion.com/
- http://en.wikipedia.org/wiki/Kinect
- http://en.wikipedia.org/wiki/Wired_glove
- http://theperegrine.com/
- http://en.wikipedia.org/wiki/Steve_Mann

C. FAQ on SmartNav

*1) Is SmartNav dangerous to damaged eyes?:* No. The explanation can be found on http://forum.naturalpoint.com/forum/ubbthreads.php?ubb=showflat&Number=24894#Post24894 and is reproduced below:

*SmartNav operates by tracking reflected or emitted IR light that is imaged by a CMOS sensor. The sensor and emitters are tuned to 880nm, slightly above the visible spectrum, you can see them emit a slight glow when the room lights are off, and this is the very upper end of the red spectrum. The sensor and IR LEDs are covered by a very special plastic that we custom*

[12] http://live.leapmotion.com/about.html
[13] http://marcansoft.com/blog/2010/11/openlase-open-realtime-laser-graphics/

| Cameras | Lights | Gaze Info | Head pose | Calibration | Accuracy (deg) | References | Comments |
|---|---|---|---|---|---|---|---|
| 1 | 0 | PoR | — | — | $2-4$ | [47], [46], [157] | web-camera |
| 1 | 0 | LoG/LoS | — | Fully | $1-2$ | [151], [144], [145] | |
| 1 | 0 | LoG | $\approx$ | — | $<1$ | [79] | $*a$ |
| 1 | 1 | PoR | — | — | 1-2 | [103], [156], [70] | $*b$ |
| 1 | 2 | PoR | ✓ | Fully | $1-3$ | [105], [100], [43] | |
| 1+1 | 1 | PoR | ✓ | Fully | 3 | [112] | Mirrors |
| 1(+1) | 4 | PoR | ✓ | — | $<1-2.5$ | [164], [20] | |
| 2 | 0 | PoR | ✓ | — | 1 | [109] | $*c$ |
| 2+1 | 1 | LoG | ✓ | — | 0.7-1 | [135] | pan/tilt |
| 2+2 | 2 | PoR | ✓ | Fully | 0.6 | [8] | Mirrors |
| 2 | 2(3) | PoR | ✓ | Fully | $<1-2$ | [128], [127] | $*d$ |
| 3 | 2 | PoR | ✓ | Fully | — | [139][11] | |
| 1 | 1 | PoR | — | — | 0.5-1.5 | [6], [133], [136], [160] | $*e$ |

Fig. 8. Comparison of gaze estimation methods with respective prerequisites and reported accuracies. Source: [3]

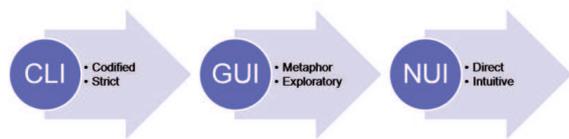

Fig. 5. The evolution of computer interfaces. Command line interfaces, graphical user interfaces and natural user interfaces. Source: Wikipedia.

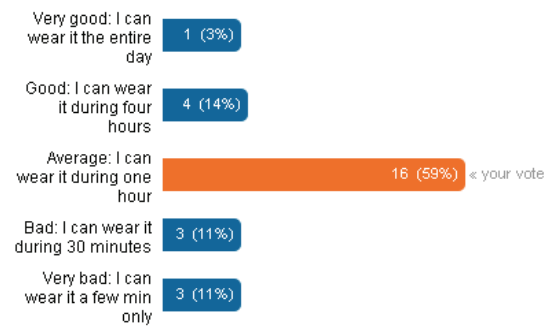

Fig. 7. Poll: "How comfortable is the NeuroSky MindWave?". Source: we initiated the poll on NeuroSky LinkedIn group, which got deleted for unknown reasons.

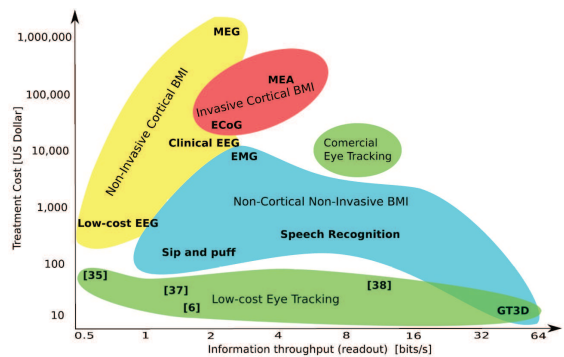

Fig. 6. Comparison of different BMI and eye tracking technologies in terms of their treatment and hardware costs (in USD) and readout performance (measured as bits/s). Source: [2]

designed with Bayer to block all light below 820nm, it passes all light above this point, it is called a band pass filter.

The LEDs emit at 880nm and are standard off the shelf IR LEDs; we run them all the time when the unit is turned on. There are 4 of them and they each have a total radiant output of about 23mw/sr, which is 23miliwats per ster radian. Total output power is NOT 4 X 23 mw/sr as the LEDS do not overlap exactly; they create a coverage pattern with slight overlap at the edges. Also, the LEDs to not emit a uniform brightness, they have an angle to half intensity, so the center of the overlapping LEDs is the SAME brightness as the center of each LEDs output, hope that makes sense.

Your eyes are sensitive to IR light, you can't see it, but your eyes will register the "power" of the light, your pupils will shrink down as if you were looking at light in the visible spectrum. Remember, we are just slightly above red in the visible spectrum. You won't feel your pupils getting smaller when our device turns on because we are a relatively low level of light for an average room condition. If you turn out all the lights in the room, put the unit about 1 foot away from your face and watch your eyes in a mirror, you will see your pupils contract, they are "seeing" the IR light.

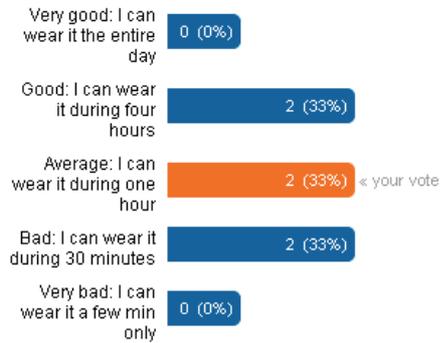

Fig. 9. Poll: "How comfortable is the Emotiv EPOC neuroheadset?". Source: we initiated the poll on Emotiv LinkedIn group, which got deleted for unknown reasons.

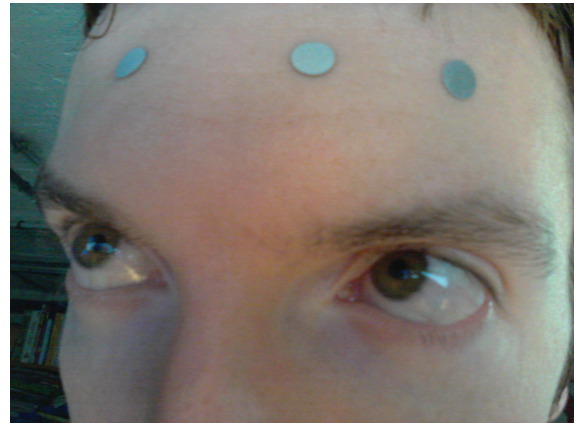

Fig. 11. Placing several reflective dots to support multiple monitors. Two dots should be enough for most configurations.

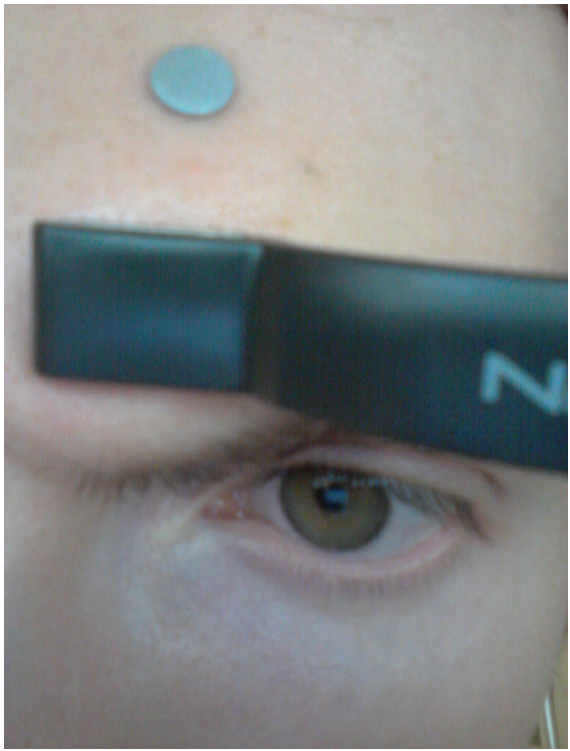

Fig. 10. Zoom on the NeuroSky and the reflective dot for SmartNav.

As for the amount of power the LEDs output, it is many of times lower than simply going outside, not to mention on a bright sunny day. As I had seen posted before, we are a small fraction of the IR output from a normal incandescent light bulb. ANSI references spec ANSI Z 136.1 - 2000 for laser power emission, but we are not a laser, so in the back of the spec they reference ANSI/IESNA RP-27.1-96, which is the spec for lamp output, basically what we are and what ANSI says to use. Maximum exposure for our wavelength range, which is from 700nm to 1100nm is 10mw/cm2. To convert our power output, which is about 30mw/sr, we apply sr x 1cm2/distance2. Typical user distance is 18" or about 45cm (on the conservative side, most users are further away), so 30mw/2025 = .015mw/cm2. Needless to say, we are on the safe side!

*2) Can SmartNav cause pain in the neck?:* No, unless SmartNav is used incorrectly. Advice on how to use SmartNav without getting pain in the neck can be found on the official forum[14].

*3) Can SmartNav be used with multiple monitors possible?:* Yes, SmartNav can be used with multiple monitors possible. The author of this article has been using SmartNav with a 6-monitor setup for several years. Figure 11 demonstrates how to place several reflective dots to support multiple monitors.

D. Eye tracking

The democratization of eye tracking is coming soon. Here are a few links:

- http://en.wikipedia.org/wiki/Eye_tracking
- http://www.tobii.com: Tobii. Over 10,000 USD.
- http://www.eyetechds.com: Eye Tech Digital Systems. Over 6,000 USD. The screen size can be up to 76cm.
- http://www.gazegroup.org/downloads: ITU GazeGroup. Free and open-source.
- http://www.cybernet.com/products/navigaze.html: NaviGaze. Free but outdated.

[14]http://forum.naturalpoint.com/forum/ubbthreads.php?ubb=showflat&Number=2619#Post2619

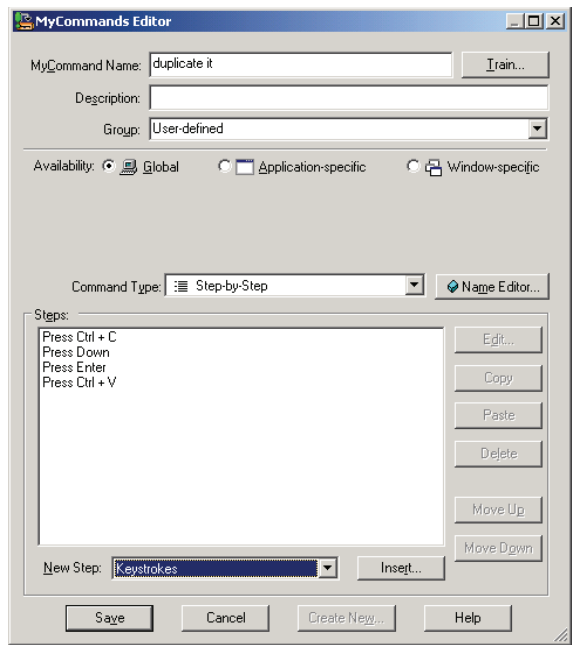

Fig. 12. Dragon NaturallySpeaking's command editor

The article [3] gives an excellent overview of the eye tracking research field. Here is the abstract: "*Despite active research and significant progress in the last 30 years, eye detection and tracking remains challenging due to the individuality of eyes, occlusion, variability in scale, location, and light conditions. Data on eye location and details of eye movements have numerous applications and are essential in face detection, biometric identification, and particular human-computer interaction tasks. This paper reviews current progress and state of the art in video-based eye detection and tracking in order to identify promising techniques as well as issues to be further addressed. We present a detailed review of recent eye models and techniques for eye detection and tracking. We also survey methods for gaze estimation and compare them based on their geometric properties and reported accuracies. This review shows that, despite their apparent simplicity, the development of a general eye detection technique involves addressing many challenges, requires further theoretical developments, and is consequently of interest to many other domains problems in computer vision and beyond.*"

### E. Brain-computer interaction

Brain-computer interaction is the ultimate human-computer interaction. We are still a long way to go but brain-computer interfaces are bound to flourish one day in a not-so-far future.

*1) NeuroSky MindWave:* Eye blink detection test with the NeuroSky MindWave:

- http://youtu.be/VoSrdnVMtz8
- http://youtu.be/XMmH0qPZqFc

*2) Emotiv EPOC neuroheadset:* http://emotiv.com/forum/forum4/topic2251/

Can wearing the Emotiv EPOC neuroheadset all the time damage skin or hair in the long term?

*No. The electrolyte is contact lens solution, which is hypoallergenic and includes non-irritant anti-microbial agents. The only material in contact with your skin is the polyester felt which is also rated for long tern skin exposure. Some people with larger heads report some discomfort from the headband pressure after an hour or two of wear. This subsides as the headset stretches a bit but it's a consideration for sure. Most people don't get any problems The saline evaporates after a few hours - you can replenish it occasionally, or you can add some glycerin to the solvent which makes it last quite a bit longer*

*3) Links:* Interesting links on brain-computer interaction:

- http://en.wikipedia.org/wiki/Brain-computer_interface
- http://en.wikipedia.org/wiki/Comparison_of_consumer_brain%E2%80%93computer_interfaces
- http://openeeg.sourceforge.net/

### F. Dragon NaturallySpeaking and speech recognition

The speech recognition software Dragon NaturallySpeaking (Windows/Mac OS, not Linux) offers the most accurate recognition amongst the programs for desktop computers. It has many more features beyond speech recognition. Defining one's own commands can be extremely powerful while being accessible for everyone, as shown in figure 12. Below is a list of websites that contain useful advice on how to optimize the use of Dragon NaturallySpeaking, in addition to the Dragon NaturallySpeaking user guide:

- http://speakeasysolutions.com
- http://www.pcspeak.com/hints/
- http://www.knowbrainer.com
- http://www.emicrophones.com/t-links_articles.aspx
- http://speechrecsolutions.com/accuracy.htm
- http://www.emicrophones.com/t-keysteps.aspx
- http://speechempoweredcomputing.co.uk/Newsletter/?p=4

Forums:

- http://www.knowbrainer.com/forums/forum/index.cfm
- http://www.speechcomputing.com/forum